\input{epsf}

\documentclass[preprint,preprintnumbers,amsmath,amssymb]{revtex4}

\usepackage{graphicx}

\begin{document}
\draft

\title{Reconciliation of MOND and Dark Matter theory}

\author{Man Ho Chan}
\address{Department of Physics, The Chinese University of Hong Kong 
\\ Shatin, N.T., Hong Kong}

\begin{abstract}
I show that Modified Newtonian Dynamics (MOND) is equivalent to 
assuming an isothermal dark matter density profile, with its density 
related to the enclosed total baryonic mass. This density profile can be 
deduced by physical laws if a dark matter core exists and if the baryonic 
component is spherically-symmetric, isotropic and isothermal. All the 
usual predictions of MOND, as well as the universal constant $a_0$, can 
be derived in this model. Since the effects of baryonic matter are larger 
in galaxies than in galaxy clusters, this result may explain why MOND 
appears to work well for galaxies but poorly for clusters. As a 
consequence of the results presented here, MOND can be regarded as a 
misinterpretation of a particular dark matter density profile.
\end{abstract}
\maketitle

\subsection{Introduction}
The dark matter problem is one of the key issues in modern astrophysics 
(cold dark matter (CDM) particles being the generally 
accepted model). The CDM model 
can provide excellent fits on large-scale structure observations 
such as the Ly$\alpha$ spectrum \cite{Croft,Spergel}, the 2dF Galaxy 
Redshift Survey \cite{Peacock} and 
cosmic microwave background \cite{Spergel2}. However, no 
CDM particles have been detected directly. Besides, the CDM model also 
encounters many well-known unresolved issues. For example, the results of 
numerical N-body 
simulations based on the CDM theory predict that the density profile of 
dark matter (the NFW profile) should be singular at the center 
\cite{Navarro} while observations in 
dwarf galaxies indicate the contrary \cite{Salucci,Borriello,Oh}. This 
problem with CDM is known as 
the core-cusp problem \cite{deBlok}. Another problem with CDM is that 
computer simulations 
predict that there should exist thousands of small dark halos or dwarf 
galaxies in the Local Group while the observations only reveal less than 
one hundred such galaxies \cite{Moore,Spergel,Cho}. This discrepancy is 
known as the missing satellite 
problem. Recent studies on the baryonic effects suggest that supernova 
feedback or radiation pressure of massive stars may provide a solution to 
this problems \cite{deBlok,Weinberg,Maccio}. Alternatively, the problems 
can also be solved if the dark matter particles are weakly interacting or 
warm \cite{Spergel,Vogelsberger,Cho}.

An alternative theory to dark matter uses the 
Modified Newtonian Dynamics (MOND) - a modification of Newton's second 
law in the weak acceleration limit \cite{Milgrom,Sanders}. 
It is suggested that a wide range of observational data including the 
rotation curves of galaxies and the Tully-Fisher relation are consistent 
with the MOND's predictions but not for CDM model \cite{Sanders, McGaugh}. 
However, recent analyses claim that it would be oversimplified 
to falsify the CDM paradigm based on such data 
\cite{Foreman,Desmond,Dutton}. Moreover, recent 
data from gravitational lensing and hot gas in clusters 
challenge the original idea of MOND without any dark matter (classical 
MOND) \cite{Angus,Sanders,Natarajan}. Thus, Sanders (1999) studied 93 
X-ray emitting clusters and 
pointed out that the missing mass still exists in cluster dark matter 
\cite{Sanders2}. 
Later, studies of gravitational lensing and hot gas 
in clusters show that the existence of 2 eV active neutrinos - the current 
upper limit of the active electron neutrino mass - is still not enough to 
explain the 
missing mass in clusters. Some more massive 
dark matter particles (e.g. sterile neutrinos) are required to account for 
the missing mass \cite{Natarajan,Takahashi,Angus2,Angus3}. In addition, 
the observational data from the Bullet cluster and the Cosmic 
Microwave Background (CMB) indicate a large amount of dark matter is 
needed to explain the lensing result and the CMB spectral shape 
respectively \cite{Clowe,Dodelson,Spergel2}.  
Another big challenge facing MOND 
is the observed shape of the matter power spectrum, which does not match 
the prediction from MOND \cite{Dodelson}. Besides these numerous big 
problems, a long list of fundamental physics difficulties such as 
violating the conservation of momentum exist in MOND 
theory \cite{Scott}. The relativistic version of MOND theory (TeVeS) is 
also not supported by recent gravitational lensing results in clusters 
\cite{Ferreras}. In summary, the observations at small scales 
may 
favor the classical MOND theory, but there are many conceptual problems 
and discrepancies in large scale observations.

Previously, Kaplinghat and Turner(2002) suggested that the MOND theory may 
be just a misleading coincidence. They show that the 
Milgrom's law - i.e., that the
gravitational effect of dark matter in galaxies only becomes important 
where accelerations are less than about $10^{-8}$ cm s$^{-2}$, can be 
explained from a cosmological cold dark matter model \cite{Kaplinghat}. 
This suggests that the 
MOND theory may be just another equivalent form of dark matter theory. 
Later, Dunkel(2004) shows that the generalized MOND equation can be 
derived from Newtonian dynamics for some specified dark matter 
contribution \cite{Dunkel}. Besides, for a suitably chosen interaction 
between dark matter, baryons and gravity, the cold dark matter model and 
MOND appear in different physical regimes of the same theory 
\cite{Bruneton}. The above studies suggest that MOND is probably not a new 
theory, but only a coincidence. In fact, most of the apparent 
successes are related to flat rotation curves in galaxies, which can also 
be explained in dark matter model. Follow from this idea, I will show in 
another way that MOND is equivalent to a particular specified type of dark 
matter density profile. This profile can be derived exactly from 
existing physical laws. The key equation and the universal constant $a_0$ 
suggested by MOND can 
also be derived from this specified profile. This claim also gives an 
explanation on why MOND apparently
works well at small scales only and it supports the standard dark matter 
model in cosmology.

\subsection{The predictions from MOND theory}
The apparent gravitation in MOND is given by
\begin{equation}
g=\sqrt{g_Na_0}
\end{equation}
when $g_N \ll a_0$, where $g_N=GM_B/r^2$ is the Newtonian gravitation 
without dark matter, $M_B$ is the enclosed baryonic mass and $a_0 
\approx 10^8$ cm s$^{-2}$ is a constant \cite{Sanders}. Generally, this 
simple form gives 
4 important predictions in galaxies and clusters without the need for dark 
matter. First, the rotational speeds of stars in a galaxy at large 
radius is given by \cite{Sanders}
\begin{equation}
v^4=GM_Ba_0.
\end{equation}
If baryons are mainly 
concentrated at the central part of a galaxy, then $M_B$ is nearly a 
constant which gives flat rotation curves \cite{Sofue}. 
Also, this equation represents the
baryonic Tully-Fisher relation $M_B \propto v^4$. The power-law dependence 
and the proportionality constant $(Ga_0)^{-1} \approx 75M_{\odot}$ 
km$^{-4}$ s$^{4}$ generally agree with the fits from observations 
$M_B=[(47 \pm 6)M_{\odot}$ km$^{-4}$ s$^4]~v^4$ \cite{McGaugh}. Secondly, 
there 
exists a critical surface density $\Sigma_m \approx a_0/\pi G$ such that 
there should be a 
large discrepancy between the visible and dynamical mass when the surface 
density $\Sigma \le \Sigma_m$ \cite{Sanders}. That means the apparent 
dark matter content 
is larger in the low surface brightness (LSB) galaxies. Moreover, MOND 
predicts that the rotation curves in LSB galaxies would continuously rise 
to the final asymptotic value \cite{Sanders}. These predictions have 
been verified 
by observations \cite{Casertano,Sanders}. Thirdly, since dark matter 
doesn't exist in MOND, the feature of 
rotation curves can be traced back to the features in the baryon mass 
distribution \cite{Milgrom2}. This prediction is generally supported by 
the rotation 
curves observed in galaxies \cite{Sanders}. Lastly, the 
dynamical mass in a 
cluster at large radius predicted by MOND is given by 
\cite{Sanders3}
\begin{equation}
M_{dyn}=(Ga_0)^{-1} \left(\frac{kT}{m} \right)^2 \left( \frac{d \ln 
\rho}{d \ln r} \right)^2,
\end{equation}
where $T$, $m$ are the mean temperature and mass of a hot gas particle and 
$\rho$ is the density profile of the hot gas. As no dark matter 
is present in clusters, MOND predicts $M_B=M_{dyn}$. However, the 
observed 
hot gas mass does not match the predicted dynamical mass 
even active neutrinos are taken into account 
\cite{Sanders2,Natarajan,Angus2}. 
Moreover, since 
$d \ln \rho/d \ln r$ is nearly a constant at large $r$ \cite{King}, we 
have $M_{dyn} 
\propto T^2$. Recent observations from 118 clusters indicate that $M_{dyn} 
\propto 
T^{1.57 \pm 0.06}$ \cite{Ventimiglia}, which shows a large discrepancy 
with the MOND's prediction.  

\subsection{Equivalent dark matter density profile of MOND}
All the above predictions can be deduced from Eq.~(1) - the only key 
equation in MOND. In the following, I will show that the above key 
equation in MOND and the 
universal constant $a_0$ can indeed be deduced by existing physical laws 
and some special properties of the dark matter distribution.

Assume that the baryonic component in a galaxy is 
spherically-symmetric and isotropic. Since the mean 
free path of the baryonic matter is small ($\lambda \sim 0.001$ pc), the 
collision between the baryonic matter is vigorous. If the interaction 
among baryons is larger than the gravitational interaction between the 
baryons and dark matter, the 
baryonic distribution would become isothermal and provide a feedback to 
the dark matter distribution. This can be justified by observations in 
many galaxies \cite{Evans}. The effect of gravity by 
baryonic component can be analysed by using the steady-state Jeans 
equation \cite{Evans}
\begin{equation}
\frac{d(\rho_B \sigma^2)}{dr}=- \rho_B \frac{d \psi}{dr},
\end{equation}
where $\rho_B$ is the baryonic mass density, $\sigma$ is the velocity 
dispersion of baryonic matter and $\psi$ is the total gravitational 
potential (includes the baryonic matter and dark matter). Since the 
isothermal distribution of baryons 
corresponds to the constant velocity dispersion $\sigma$, by Eq.~(4), we 
get \cite{Evans}
\begin{equation}
\sigma^2 \frac{d \rho_B}{d \psi}+ \rho_B=0.
\end{equation}
The solution to the above equation is $\psi=\psi_0- \sigma^2 \ln \rho_B$, 
where $\psi_0$ is a constant. Substituting the function $\psi$ into the 
Poisson equation and assuming the total mass is dominated by dark matter, 
the dark matter density is given by
\begin{equation}
\rho_D=- \frac{\sigma^2}{4\pi G} \left[ \frac{1}{r^2} \frac{d}{dr} 
\left(r^2 \frac{d \ln \rho_B}{dr} \right) \right].
\end{equation}
Let $\gamma=- d \ln \rho_B/d \ln r$, we get
\begin{equation}
\frac{\gamma}{r^2}+ \frac{1}{r} \frac{d \gamma}{dr}= \frac{4 \pi G 
\rho_D}{\sigma^2}.
\end{equation}
Since the isothermal baryonic component gives $\gamma=2$, we get
\begin{equation}
\rho_D= \frac{\sigma^2}{2 \pi Gr^2}.
\end{equation}  
Nevertheless, observational 
data in galaxies strongly support the existence of a core in the dark 
matter 
density profile \cite{deBlok}. The existence of a small core (size $\sim$ 
kpc) may be due to the self-interaction between dark matter particles 
\cite{Vogelsberger} or the baryonic processes such as supernova feedback 
\cite{deBlok}. Therefore, we may slightly modify Eq.~(8) without 
destroying the isothermal distribution at 
large radius by introducing a cored-isothermal profile:
\begin{equation}
\rho_D= \frac{\sigma^2}{2 \pi G(r^2+r_c^2)}= \frac{\rho_c}{1+(r/r_c)^2},
\end{equation}
where $\rho_c$ and $r_c$ are the central density and core radius of the 
dark matter profile respectively. When $r \gg r_c$, Eq.~(9) will reduce 
to Eq.~(8). 

Recent observations in galaxies indicate that the product of the dark 
matter central density and core radius is a constant: 
$\rho_cr_c=141^{+82}_{-52}M_{\odot}~ {\rm pc}^{-2}=C$ \cite{Gentile}. 
By using the profile in Eq.~(9), we get
\begin{equation}
\frac{\sigma^2}{2 \pi Gr_c} \approx C.
\end{equation}
The total enclosed dark matter mass within $r_c$ is $M_c \approx  
\sigma^2r_c/G$. Therefore we have
\begin{equation}
\frac{\sigma^4}{2 \pi G^2M_c} \approx C.
\end{equation}
Furthermore, since $M_c \sim 0.1M_{DM}$ \cite{Rocha} and $M_B \approx 
0.2M_{DM}$ \cite{Spergel2}, we have
\begin{equation}
\sigma^4 \sim \pi G^2CM_B.
\end{equation}
As the MOND effect is important at large radius only, by substituting 
Eq.~(12) into 
Eq.~(8), the resulting dark matter density profile is given by
\begin{equation}
\rho_D \approx \frac{1}{4 \pi r^2} \sqrt{4 \pi CM_B}= \frac{1}{4 \pi 
r^2} \sqrt{\frac{M_Ba_0}{G}},
\end{equation}
where we assume $a_0=4 \pi CG \sim 10^{-8}$ cm s$^{-2}$. Since the 
baryonic mass $M_B$ is nearly a constant at large radius, the 
total mass for dark matter is
\begin{equation}
M_{DM}=4 \pi \int_0^r \rho_Dr^2dr \approx \sqrt{ \frac{M_Ba_0}{G}}r.
\end{equation}
If the apparent gravity $g$ in MOND is indeed a real gravitational effect 
from dark matter, by using the above equation, we get 
\begin{equation}
g=\frac{GM_{DM}}{r^2}=\sqrt{\frac{GM_Ba_0}{r^2}}=\sqrt{g_Na_0},
\end{equation}
which is the same key equation in MOND theory. The corresponding constant 
$a_0$ surprisingly matches the universal constant suggested in MOND: $a_0 
\approx (1.3 \pm 0.3) \times 10^{-8}$ cm s $^{-2}$ \cite{McGaugh}. In 
other words, our result suggests that the basic assumption in MOND theory 
(Eq.~(1)) is equivalent to the specified dark matter profile in Eq.~(13). 
Therefore, most of the predictions by MOND theory can also be obtained 
by our specified dark matter profile. For example, the baryonic 
Tully-Fisher relation can be obtained in Eq.~(12). Moreover, since 
$M_{DM}$ 
varies with $M_B$ (see Eq.~(14)), the rotational speed $v$ also varies 
with $M_B$. Therefore, tiny variation in baryonic mass distribution can 
directly reflect on the rotation curve. This result generally matches the 
third MOND's prediction. Therefore, the apparent success of MOND is 
telling us that the dark matter density distribution is related to the 
baryonic matter content $M_B$ and the velocity dispersion of dark matter 
particles is nearly uniform. These properties can be 
derived from existing physical laws.

\subsection{Discussion}

Traditionally, the dark matter problem has been mainly addressed by the 
existence of cold dark matter. However, the successful 
predictions from MOND on the galactic scale may indicate that MOND is 
correct to a certain extent. Generally, these two theories are highly 
incompatible. In this article, I show that the basic assumption in MOND is 
equivalent to a particular form of dark matter density profile. This form 
can be naturally obtained if the distribution of the baryonic matter is 
spherically-symmetric, isotropic and isothermal. Also, empirical 
studies show that $\rho_cr_c$ is a constant for most galaxies. This 
relation can be derived in some particular models of self-interacting dark 
matter \cite{Chan}. By using these two properties, the 
derived dark matter density profile is equivalent to that in MOND 
theory, and all predictions from MOND can be obtained. The 
universal constant $a_0$ suggested in MOND can also be derived in this 
model: $a_0=4 \pi CG \sim 10^{-8}$ cm s$^{-2}$, which gives excellent 
agreement with MOND's prediction. In fact, the isothermal dark matter 
density profile in galaxies is well supported by many recent observations 
\cite{Ven,Velander,Westfall}. Moreover, since the charateristics of 
cores and the effect of baryonic matter are mainly found in galaxies, this 
result also gives an explanation why MOND apparently works well in 
galaxies only. 

In fact, several severe challanges facing MOND such as the missing mass in 
clusters, the shapes of matter power spectrum and CMB spectrum indicate 
that MOND probably is not a universal law in physics. If MOND indeed 
represents an isothermal distribution of dark matter that only works in 
galaxies but not in clusters, it would not be necessary for us to make any 
changes in Newtonian dynamics as well as General Relativity. 
Therefore, the apparent sucess of MOND may be just a coincidence. However, 
apart from 
the context of dark matter, MOND also predicts that some strange effects 
may be observable in the Solar system \cite{Galianni}. For example, around 
each equinox date, 2 spots emerge on the 
Earth where static bodies experience spontneous acceleration due to the 
possible violation of Newton's second law \cite{Ignatiev}. Since these 
effects are independent of dark matter, MOND will still be survived if 
these effects can really be detected in the future.  

To conclude, the reconciliation of the MOND and dark matter model 
suggests that MOND theory is equivalent to the 
isothermal dark matter density profile in dark matter model. Also, the 
dark matter density profile on the galactic scale is related to the 
baryonic mass content. 

I am grateful to Prof. R. Ehrlich and the referees for helpful comments on 
the manuscript.

\end{document}